\documentclass[%
superscriptaddress,
twocolumn,
 amsmath,amssymb,
 aps,
pre,
]{revtex4-1}

\usepackage{graphicx}
\usepackage{dcolumn}
\usepackage{bm}

\begin{document}


\title{Dilute concentrations of submicron particles do not alter the brittle fracture of polyacrylamide hydrogels}

\author{Albert Taureg}
\author{John M. Kolinski}%
 \email{john.kolinski@epfl.ch}
\affiliation{%
\'{E}cole Polytechnique F\'{e}d\'{e}rale de Lausanne, 1015 Lausanne, Switzerland
}%

%
%

\date{\today}

\begin{abstract}
In studies of the dynamic failure of brittle hydrogels, a bound has been placed on the process zone scale - the scale where material separation and ultimate failure occur. For the polyacrylamide hydrogel system under study, this bound is set at 20 microns. Thus, any subtle alterations to the material at a \emph{smaller} scale should not in principle alter the dynamic fracture response of the hydrogel. Here we test this directly by embedding sub-micron-scale latex polystyrene microspheres within the brittle polyacrylamide hydrogel at a solids fraction of 0.1 \%. We verify that the spheres are well-distributed throughout the hydrogel material at this concentration with optical microscopy, and reconstruct the 3D distribution of these spheres using laser scanning confocal microscopy in backscatter mode. Finally, we test the fracture behavior of this gel with the dilute, embedded sub-micron spheres, and find that the brittle material failure modality common to this material \emph{without} the sub-micron spheres is indeed retained. By comparing the crack tip opening displacement, fracture energy and the crack's speed with established data from prior experimental work, we demonstrate that this material's failure is brittle, as it is in good agreement with the pure hydrogel system.
\end{abstract}

\maketitle

Soft materials are increasingly employed in manifold applications, from soft electronics\cite{dickey_stretchable_2017} to soft robotics\cite{ilievski_soft_2011}, and even drug delivery systems\cite{jeong_biodegradable_1997}. Despite the increasing importance and applicability of soft materials in our daily lives, we still don't have a clear view of what their performance limits are - indeed, recent developments of compound gel systems show that hydrogels, at 90\% water, can be made to undergo stretches exceeding ten\cite{gong_doublenetwork_2003}, exhibit recovery of their mechanical properties after extreme loading conditions\cite{sun_highly_2012, liu_tough_2017} and be facilely fabricated using a rich variety of chemistries that impart on such materials sensitivity to e.g. loading rate\cite{liu_tough_2017}. While a substantial volume of work focusing on brittle hydrogel systems has brought to light significant and important insights into the fracture behavior of brittle polyacrylamide hydrogels\cite{tanaka_discontinuous_1998, tanaka_fracture_2000, livne_universality_2005, livne_breakdown_2008, livne_near-tip_2010, goldman_acquisition_2010, goldman_intrinsic_2012, kolvin_crack_2015, kolvin_nonlinear_2017, kolvin_topological_2017}, key questions concerning crack stability and three-dimensional crack propagation remain unanswered. One of the key hindrances to the advancement of the study of 3D fracture is a general inability to resolve the deformation field very near the crack tip in 3D. Despite the many applications of particle tracking in high-resolution imaging of deformation in soft solids\cite{franck_three-dimensional_2007, bar-kochba_fast_2015}, this method requires high particle seeding densities, which might alter fracture behaviors\cite{steinhardt_rules_2019}. It is thus critical to evaluate whether at these concentrations the particles alter the fracture behavior, or behave as passive tracers. If the brittle fracture behavior is preserved at these seeding densities, this method will enable direct observation of the process zone in a brittle hydrogel, or facilitate measurement of the complex 3D deformations in-situ for a propagating crack. 

\begin{figure}[!ht]
	\includegraphics[width=\columnwidth]{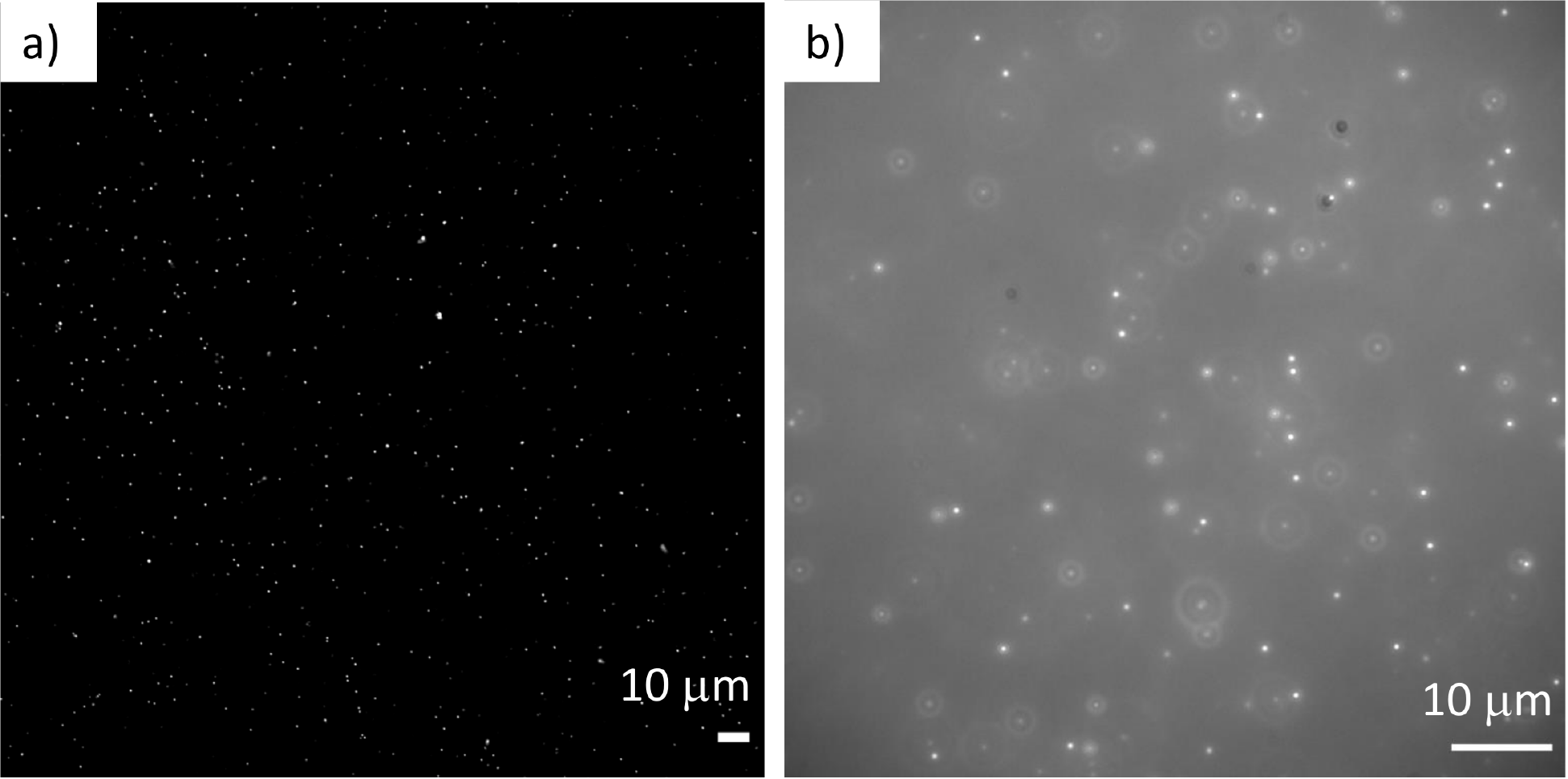}
	\caption{\label{fig1} Images of particle loading with 0.47 micron latex polystyrene spheres at 0.1\% solids fraction in the brittle hydrogel matrix. a) Darkfield imaging with 40x air-immersion objective (Nikon plan fluor, NA = 0.75) shows that in the narrow focal plane of the objective, we have a large number of particles - corresponding to a high numerical density - as expected for the micro-sphere's small diameter. b) Brightfield imaging with the same sample at a different location verifies the high numerical density of the microspheres with a 100 x objective (Nikon plan fluor oil immersion, NA = 1.3). }
\end{figure}

\begin{figure*}[!ht]
	\includegraphics[width=\textwidth]{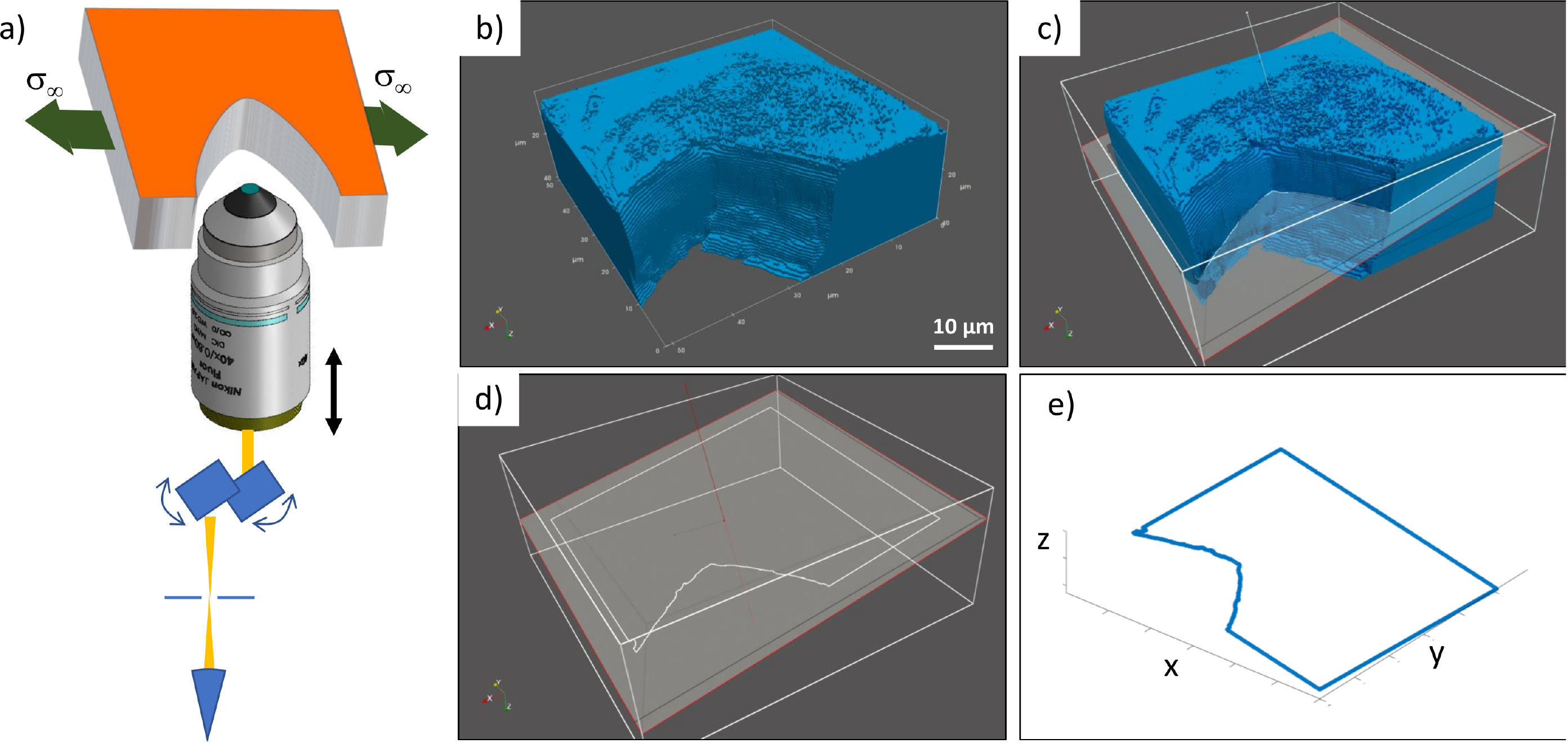}
	\caption{\label{fig2} Laser-scanning confocal micrograph of a gel sample with fluorescently labeled solvent and CTOD analysis workflow. a) A schematic showing a sample loaded in tension with our custom microscope mounted loading apparatus. The objective is mounted to a piezo-scanning mount (PI Fast PIFOC Z-Drive, 400 micron stroke) indicated by the black arrow. The fluorescence emitted within the objective's numerical aperture is de-scanned using a resonant-galvo pair (Cambridge Technologies), and passed through a spatial filter (Newport M-900), represented by the horizontal lines. This signal is then detected by an avalanche photodiode (Thorlabs APD430), and recorded using a customized laser scanning microscopy platform (Scanimage). b) A 3D micrograph of the fluorescently labeled solvent after image segmentation shows the high-resolution 3D data recorded with the confocal microscope. c) Using the 3D dataset, we select a plane aligned with the local conformation of the crack tip in 3D in silico. The surface coordinates are extracted (d) and processed for further analysis (e). }
\end{figure*}

Here we study the dynamic fracture behavior of a brittle hydrogel system with a dilute concentration of embedded polystyrene microspheres. The spheres used in this study are selected to have a size significantly smaller than the process zone scale in the gel material. The concentration employed, while dilute, averages to one particle per 10 $\mu$m$^3$, which is sufficiently dense to carry out particle tracking studies or digital volumetric correlation studies of material displacements. Using brightfield and darkfield optical microscopy, we verify the anticipated concentration of microspheres within the gel sample, and establish the uniformity of their distribution using laser scanning confocal microscopy (LSCM). We present a workflow for direct measurement of the microsphere positions with a precision signficantly better than the confocal volume of the microscope. Dynamic fracture studies are carried out in mode-I, in-plane tensile loading, and the crack's velocity and crack tip opening displacement (CTOD) are analyzed to assess whether the well-documented brittle material failure modality\cite{livne_universality_2005, goldman_acquisition_2010} is maintained.

\begin{figure*}[ht]
	\includegraphics[width=\textwidth]{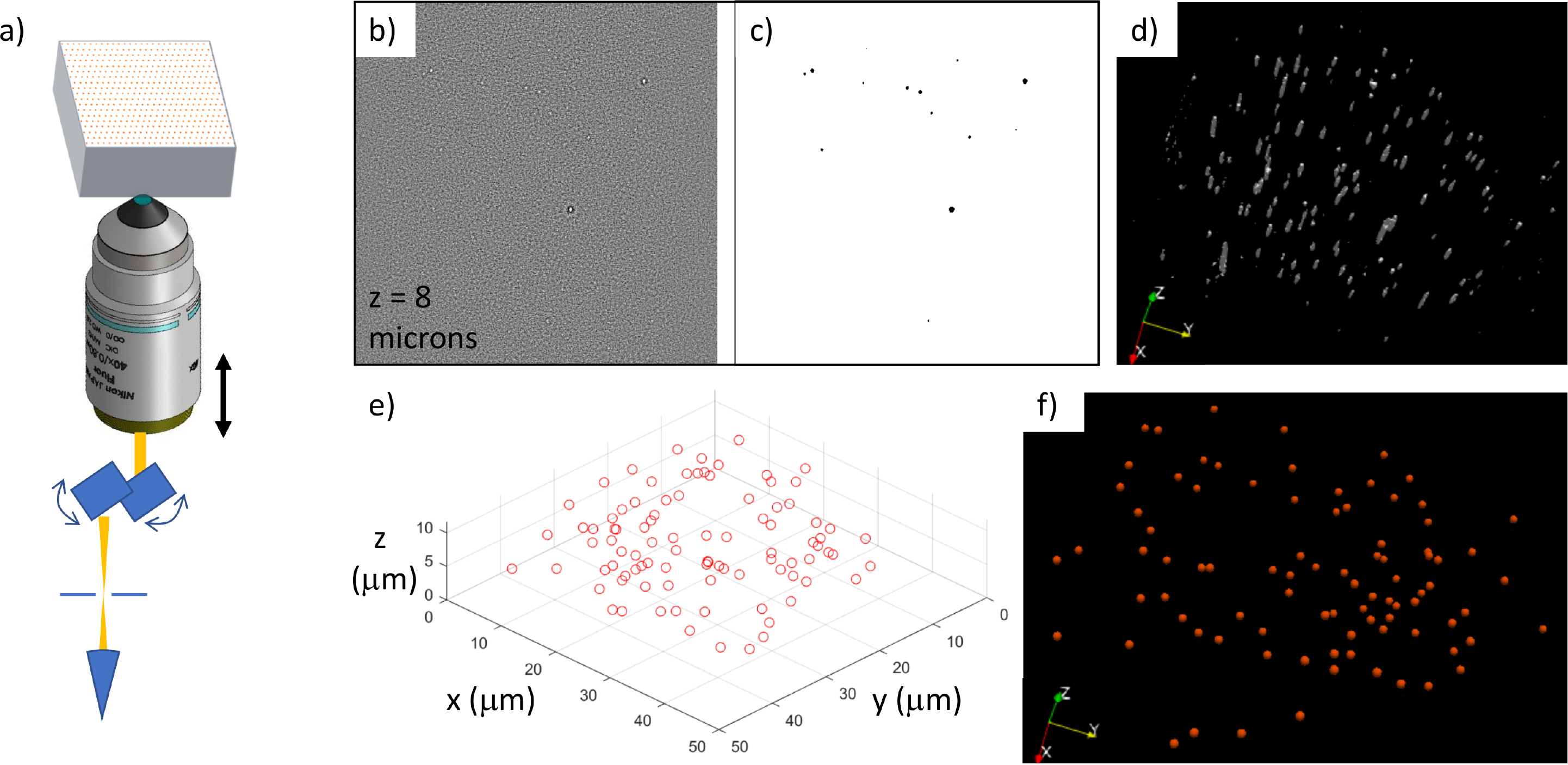}
	\caption{\label{fig3} Laser-scanning confocal micrograph of a gel sample with 0.47 micron diameter polystyrene latex spheres embedded within. a) Schematic microscope platform as described in the caption of Fig.~\ref{fig2}. b) An 8-bit slice of the z-stack recorded with the laser scanning microscope shows some bright peaks after filtering. c) Segmentation using a random forest classifier (Ilastik software\cite{berg2019}) is used to extract the positions of particles in 3D. d) The fully-segmented 3D representation of the single slice shown in (c). We see the elongation of the particle point-spread functions in $z$, as well as some noise from the imperfect segmentation scheme. e) Further filtering based on the bounding box size of the segmented particles selects only those centroids that have the anticipated point-spread function in (d). f) The particles are shown to scale in the 3D domain, showing that they are indeed dilute, and nevertheless at a high numerical density.}
\end{figure*}

Gel samples are prepared from a 13.7 \% monomer concentration (wt/vol)  with 2.7 \% bis-acrylamide cross-linker. Polymerization is initiated and catalyzed using an established protocol with Ammonium-per-sulfate and TEMED\cite{menter_acrylamide_nodate}. 0.47-micron polystyrene latex microspheres are added at a 0.1 \% solid fraction prior to the onset of polymerization. All chemistry is sourced from Sigma-Aldrich. Once the catalyst and initiator are added, the gel is allowed to polymerize between glass plates separated by 190 micron rigid plastic spacers for four hours prior to any experiments. For the dynamic fracture experiments, a 4 x 2 cm rectangle is cut from the gel sheet, and mounted on custom grips in a tensile loading apparatus, such that 1 x 2 cm area of the gel sample is confined within the gripping area on each side of the sample. A 10\% nominal strain is applied, and the crack is initiated using sharp scissors at the edge of the sample. The crack's progress is imaged onto a high-speed camera (Photron Nova S12) at 12800 fps, with an exposure time of approximately 700 nanoseconds using a long-working distance microscope (Nikon AZ100, 2x objective). A bright, collimated and pulsed LED source is synchronized with the camera's exposure to illuminate the sample in-line. The load applied to the sample is monitored with a load cell (PT Global) amplified with a lock-in amplifier (Stanford Research Systems SR865A), whose output is digitized using an National Instruments data acquisition card (NI DAQ). The displacement of the stages (Newport) used to apply the load are recorded concurrently with the load cell signal and written to a text file using a custom visual basic software to drive the experiment. 

The fractured samples are preserved for further analysis with optical profilometry. As the gel is translucent, the fracture surface is cast into PVS (Zhermack), which is then mounted to a microscope slide. This slide-mounted sample is then imaged using a Nikon interferometry objective attached to an inverted microscope platform (Nikon Ti-Eclipse). The objective scanning z-stage is used to modulate the position of the interferometry objective relative to the sample at steps of 50 nm. A weakly coherent LED light source is used to illuminate the sample, generating fringes once the working distance is precisely aligned with the reference cavity internal to the Mireau objective. A z-stack of images recorded using micro-manager software is processed using a custom MATLAB code to extract the envelope of the interference fringes, and thus identify the z-location of each point on the surface of the elastomer casting.

Quasi-static tensile fracture experiments are carried out using a custom loading apparatus configured from off-the-shelf precision stages (Newport CONCEX-MFACC) affixed to a precision 10N load cell (HBM) amplified using a ClipX amplifier (HBM). Simultaneously, the displacement of the stages is recorded using an LVDT (HBM) with a stroke of 1 cm. Digital signals are either recorded using an NI DAQ or an oscilloscope. 

The loading apparatus is mounted on a custom-built laser-scanning confocal microscope platform based on the Scanimage software platform (Vidrio Technologies LLC). All LSCM imaging is carried out using a Nikon water immersion objective (CFI APO NIR, NA = 0.8) mounted to a piezo objective scanning mount (PI Fast PIFOC Z-Drive, 400 micron stroke). Thorlabs and Newport optical components are used to guide the beam, and mount the resonant-galvo scanning mirror pair. Laser illumination is focused onto the sample from a NKT Photonics white laser system (NKT Super-K EVO) driven with an acousto-optic filter with up to eight simultaneous emission lines. Emitted light is collected after passing through a dichroic (Semrock) and a spatial filter (Newport M-900) using a Thorlabs avalanche photodiode (APD430). The signal is digitized using the Scanimage acquisition system, where it is synchronized with the scanning mirrors and the z-position in order to obtain a 3D reconstruction.


The small size of the 0.47 micron diameter microspheres enables us to obtain a high numerical density at a low solid fraction of 0.1\%. The numerical density is verified using two different optical microscopy modalities, including darkfield and brightfield imaging, as shown in Fig.~\ref{fig1} (a) and (b), respectively. We observe that the particles are of high numerical density at two arbitrarily selected locations within the sample, and that their distribution appears to be random, and devoid of aggregation within the particle-laden gel sample, as shown in Fig.~\ref{fig1}.

\begin{figure}[h!]
	\includegraphics[width=\columnwidth]{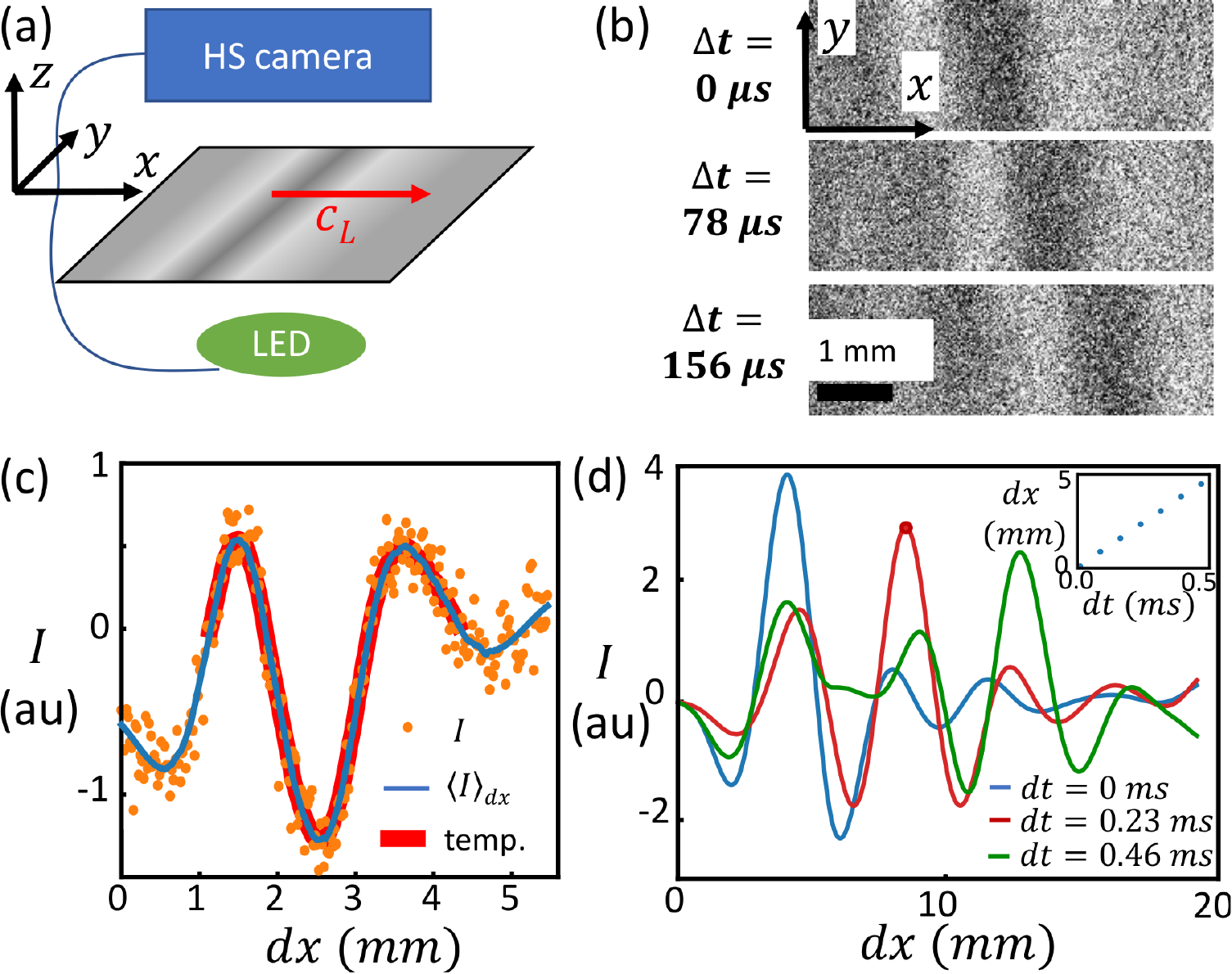}
	\caption{\label{figspeed} In-situ characterization of material properties using longitudinal wave speed measurements. a) Schematic of the dynamic wave propagation observation for in-situ characterization of the material. Collimated illumination is scattered by the deformation of the material during longitudinal wave propagation due to gradients in the $zz$-strain component. b) An image sequence of the longitudinal disturbance propagating through the sample at three time steps. Images are normalized by the preceding 10 images to ensure background uniformity. c) The intensity trace along $x$ is determined by averaging the images shown in (b) over the $y$ axis. Local smoothing using the Savitzky-Golay method serves as a low-pass filter for the data. A template of the intensity trace is extracted for further processing, as this wave envelope is conserved during wave propagation. d) Matched filtering of the intensity traces at subsequent time steps is carried out using the template extracted from the smoothed trace as shown in (c). The intensity peak corresponds to the maximum in the filtered data. To further reduce noise and enhance precision, a parabola is fitted to these data over a window of $\pm15$ pixels, or $\pm$330 $\mu$m. Finally, a linear fit is applied to the peak-location data over time; the slope corresponds to a measurement of the wave propagation velocity, shown inset.}
\end{figure}

In order to correctly resolve the positions of the particles embedded within the gel near a free interface, we must match the optical indices of the gel material and free-space to prevent the introduction of optical aberration\cite{hecht2012optics}. We verify the absence of optical aberration using an alternative 3D imaging modality, wherein we label the solvent in a gel sample \emph{without} particles, loaded in tension using the loading apparatus mounted to the LSCM; here, a crack is inserted to obtain the free-interface. A typical image is shown for reference in Fig.~\ref{fig2} (a). Using the 3D imaging capability of the LSCM, we can reconstruct the position of the free interface very precisely. Index matching between the gel sample and the `free-space' behind the crack is acheived using fluorinated oil (FC-40). The veracity of the 3D reconstruction is shown in Fig.~\ref{fig2} (b). A workflow that can be used to carry out crack tip opening displacement (CTOD) measurements with an arbitrarily oriented plane in 3D is demonstrated in Fig.~\ref{fig2} (c)-(e), where we obtain the 3D coordinates that pass through a selected plane in 3D. These have been imported into MATLAB for further processing and analysis. Visualizations are carried out using Paraview.

\begin{figure*}[ht]
	\includegraphics[width=\textwidth]{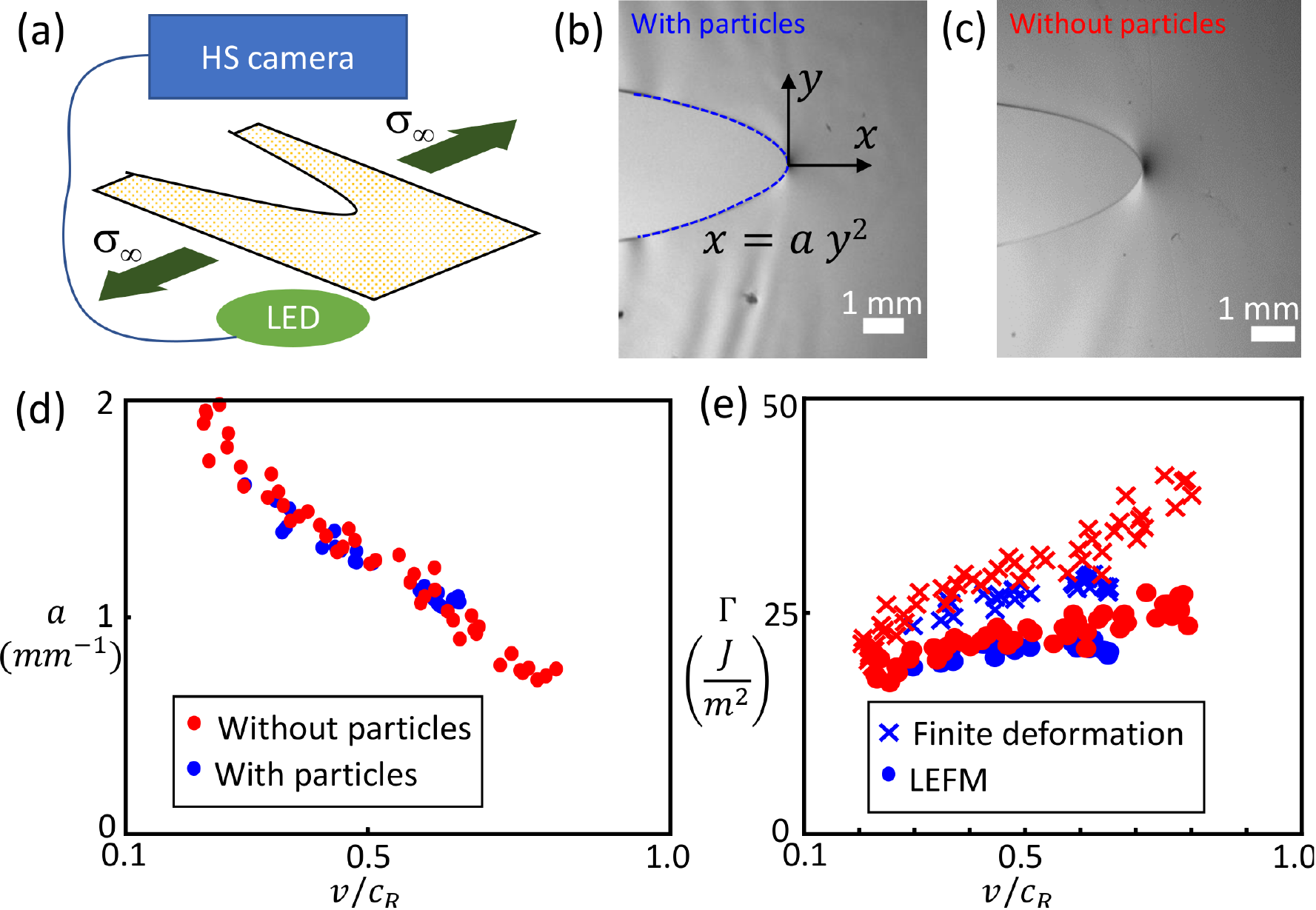}
	\caption{\label{figagamma} Dynamic fracture tests of gel material with and without 0.47 $\mu$m diameter PS spheres at 0.1\% volume fraction. a) Schematic of dynamic fracture set-up used to record plan-view images of cracks propagating through hydrogels. b) A plan-view image of a crack propagating smoothly through a hydrogel with particles at a velocity of 0.4$c_R$. Notably absent are microbranching events observed in prior work\cite{boue_origin_2015}. The parabolic region of the crack tip opening displacement (CTOD) used to extract the crack-tip curvature is depicted with the blue dashed line. c) A crack propagating at a similar velocity to that shown in (b) through the hydrogel material \emph{without} particles shows a similar smooth fracture propagation. d) The crack-tip curvature measured from the CTOD is plotted as a function of the noramlized velocity for cracks propagating in gels with (without) particles in blue (red). We see that despite slight differences in sample geometry, the CTOD curvature is comparable. e) The fracture energy $\Gamma$ is measured as a function of normalized velocity for gels with (without) particles in blue (red). The values of $\Gamma$ are determined in two ways: first, with a biaxial, finite-deformation model (x's), and second with the Linear Elastic Fracture Mechanics (LEFM) solution data (circles). }
\end{figure*}

Despite their small size, the 0.47 micron polystyrene latex spheres are easily visible using the confocal microscope. Indeed, spheres of this size have already been used successfully in displacement field studies of hydrogels with LSCM\cite{bar-kochba_fast_2015}. Here, we are simply imaging their location in space using the same optical setup as before, without the free-surface adjacent to the gel sample, as shown in Fig.~\ref{fig3} (a). A single slice of the 3D z-stack of images is shown after filtering in Fig.~\ref{fig3} (b). Small bright spots indicate the presense of a microsphere in the confocal volume of the microscope for this slice. These slices are then segmented using a random forest classifier (Ilastik software\cite{berg2019}), which enables 3D segmentation of the particle locations, as shown for the same slice in Fig.~\ref{fig3} (c). Once segmentation is carried out, we plot the point spread functions for the microsphere in 3D using Paraview visualization software, as shown in Fig.~\ref{fig3} (d). Any residual noise from the segmentation scheme can be filtered using the dimensions of the point spread functions from the data in Fig.~\ref{fig3} (d), and the centroids of these features can be readily identified using established software routines in MATLAB. The centroids of all candidate particles are shown in Fig.~\ref{fig3} (e). These centroid locations are then plotted using spherical glyphs to represent their location and size, to scale, in Paraview; this plot shows how dilute the spheres are, and furthermore how uniformly distributed in the gel they are, verifying their utility for measuring material displacements, as shown in Fig.~\ref{fig3} (f).

\begin{figure}[h!]
	\includegraphics[width=\columnwidth]{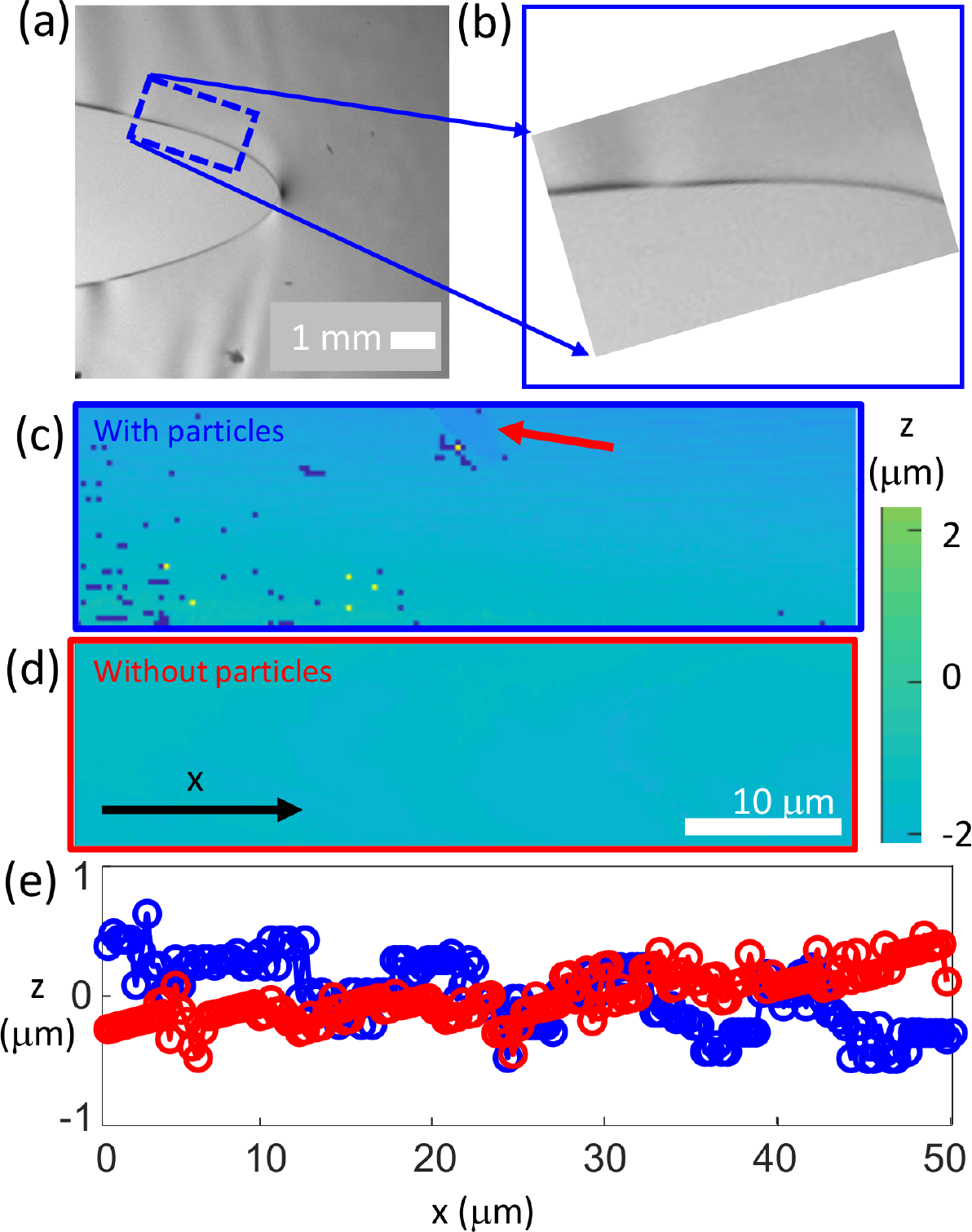}
	\caption{\label{figprof} Profilometry of fractured hydrogel crack surface. a) An in-situ, plan-view image of a crack propagating through a particle laden hydrogel. The dashed rectangle indicates a representative view of the surface later analyzed using a profilometer in the $x-z$ plane. b) A typical surface analyzed by profilometry after the fracture surface is shown zoomed-in, highlighting the absence of surface texture on the scale of a single pixel; here this is $21 \mu$m. c) Optical profilometry of the post-mortem fracture surface shows that it is very smooth on the scale of a micron. Some noise resulting from an imperfect optical reconstruction appears near the left-hand side of the image. A small feature, likely due to a particle, can be seen near the middle-top of the image; otherwise, the surface is unblemished d) The surface of the identical gel composition, fractured in an identical manner, is provided for comparison. Here, we see that the majority of the surface is as smooth as that of the gel with particles, suggesting that the microspheres have not significantly affected the crack's dynamics. e) Traces along the profile of the sample with the microspheres (blue) and without the microspheres (red) show that the surface has a comparable roughness with and without the particles. }
\end{figure}

While we ultimately must test the fracture behavior of the gel-particle system, it is imperative that we first establish the material properties of the sample to ensure our measurement of the fracture energy $\Gamma$ is accurate, as the shear modulus enters directly into the expression for $\Gamma$. To do this, we follow a method introduced in prior experimental study of planar mode-I cracks in double-network hydrogels, where a strain pulse is used to measure the material's sound speed\cite{kolvin_how_2018}. Here, we are not spray painting the gel, however, so an alternative approach is required to measure the propagation of longitudinal waves in the material, using an in-line imaging set-up with a collimated, pulsed light source and our high-speed camera, as shown schematically in Fig.~\ref{figspeed} (a). During propagation, longitudinal waves introduce a subtle lensing effect: the volumetric strain is accommodated in the $x-z$ plane, and the strain gradients in the $zz-$component diffract the collimated light, introducing modulations in the intensity at the wavefront. This appears as a fractional intensity modulation, of the order of 0.5 \% over the background. As this signal is very small, and larger than the sensor's noise, we must use matched filtering to extract the front position. This is achieved first by extracting a set of background images to normalize the intensity in the region of interest; a time-series of the normalized images during wave propagation is shown in Fig.~\ref{figspeed} (b); here, the contrast is significantly enhanced to highlight the shape of the longitudinal wave-front. As this is a plane wave, we average over the $y-$axis to further reduce the noise; these data are shown in the yellow points in Fig.~\ref{figspeed} (c). low-pass filtering over the $x-$axis further reduces noise, leading to a signal that can be processed by matched filtering via a template formed from the wavefront shape, as shown by the blue and red curves in Fig.~\ref{figspeed} (c), respectively. The template is then used with linear filtering to identify the position of the wave front from the peak in the match-filtered data; sub-pixel resolution of the front position is obtained by fitting a parabola to the peak of the filtered data, as indicated by the curves and point in Fig.~\ref{figspeed} (d). The wave speed is then extracted from the sub-pixel resolved front position's temporal evolution by fitting a line through the position trace, as shown in Fig.~\ref{figspeed} (d) inset. Once the wave speed is established, we can directly extract the material's shear modulus by recognizing that the shear wave speed $c_s$ is 1/2 of the longitudinal wave speed $c_l$ (measured), and $c_s = \sqrt{\mu/\rho} \rightarrow \mu = \rho (c_l/2)^2$. For reference, a table of measured moduli for the gel material with and without particles is included in the supplementary material. The shear modulus is on average 7\% less for the particle-laden gel than the particle-free gel.

Upon embedding the spheres within the brittle hydrogel, it is unclear whether they might alter the brittle fracture mechanics germane to this material system\cite{livne_near-tip_2010, goldman_acquisition_2010}. Indeed, recent studies suggest that particles of varying sizes can strongly affect a crack's direction and stability\cite{steinhardt_rules_2019, rosen-levy_maximum_2020}. The particles are nearly two orders of magnitude smaller than the 20 micron bound on the process zone size in this material\cite{livne_near-tip_2010}. Here we test the particle-laden hydrogel sample subject to mode-I tensile loading (experimental set-up is shown schematically in Fig.~\ref{figagamma} (a)), and find that the canonical parabolic CTOD observed during brittle fracture of such hydrogels in prior studies\cite{livne_universality_2005,  goldman_acquisition_2010} emerges, as shown in Fig.~\ref{figagamma} (b). Notably, disturbances typical of crack front instability are absent\cite{boue_origin_2015, kolvin_topological_2017}; for comparison, we include a smooth CTOD from the standard gel composition at a similar loading geometry and crack speed in Fig.~\ref{figagamma} (c). If instead the fracture behavior were affected by the 0.47 micron particles at 0.1\% solid fraction, the smooth parabolic profile would not be observed, as occurs with dilute concentrations of larger particles\cite{rosen-levy_maximum_2020}. To calibrate the material's fracture properties, we extract the crack-tip curvature for a time-series of images during the crack's acceleration, as shown in Fig.~\ref{figagamma} (d). As a function of normalized velocity, we find that the CTOD curvature agrees well for the gel material with and without particles. Using a recent theory for calculating the fracture energy $\Gamma$ from the CTOD curvature\cite{goldman_boue_failing_2015}, we find that the materials have very similar $\Gamma (v/C_R)$ curves, indicating that the brittle failure properties are not altered by the sub-micron particles at 0.1\% solid fraction, whether finite stretch is taken into account (x's) or not (circles), as shown in Fig.~\ref{figprof} (e). 

While the measurements of $\Gamma(v/c_R)$ suggest that the material's brittle failure behavior is not altered by the particles, post-mortem profilometry can provide additional insight into the detailed, small-scale processes occuring during crack propagation. A side-on image showing the surface analyzed using profilometry is shown in Fig.~\ref{figprof} (a) and (b). The fracture surface is essentially devoid of features, as shown in Fig.~\ref{figprof} (c); indeed, an identical region probed with the same profilometry method on the surface of a gel sample without the particles appears to be of comparable surface roughness, as shown for comparison in Fig.~\ref{figprof} (d). Traces along the crack surface in the $x$-direction show that the surfaces have a comparably smooth surface, as shown in Fig.~\ref{figprof} (e) at a 50 nm scale. 

We have demonstrated that dilute embedding of rigid polystyrene spheres in an otherwise brittle polyacrylamide hydrogel matrix does not alter the brittle fracture response for a dynamic crack. Indeed, the fracture energy compares well with similar values of the fracture energy in the brittle hydrogel \emph{without} the 0.47 micron diameter microspheres over a wide range of normalized crack velocity, from $v/c_R$ = 0.2 - 0.8, when the stress fields near the crack tip are significantly altered due to dynamic effects\cite{freund_dynamic_1998, fineberg_instability_1999}. 

Using a variety of microscopy modalities, We have thoroughly verified that the distribution of the microspheres is random, and that we nevertheless obtain a high numerical density, with one sphere per 10 $\mu m^3$ on average. This high particle density can be exploited to uncover material displacements very near the crack tip, provided a suitable loading platform and microscopy imaging modality are available. Indeed, our LSCM platform and custom-made loading platform are well-suited for this application, and can be used to shed insight into the failure of hydrogels very near the crack's advancing tip, all while generating precise, 3D data of either the crack tip opening displacement, or indeed, the relative displacements of the particles from reference configuration within the gel itself. These experiments will require further work, but can build upon the technical foundations validated in the images shown in Figs.~\ref{fig2} and \ref{fig3}.

\begin{acknowledgments}
	JMK and AT gratefully acknowledge the support of Innosuisse Impulse grant 35778.1.
\end{acknowledgments}

\bibliography{biblio}

\begin{thebibliography}{28}%
\makeatletter
\providecommand \@ifxundefined [1]{%
 \@ifx{#1\undefined}
}%
\providecommand \@ifnum [1]{%
 \ifnum #1\expandafter \@firstoftwo
 \else \expandafter \@secondoftwo
 \fi
}%
\providecommand \@ifx [1]{%
 \ifx #1\expandafter \@firstoftwo
 \else \expandafter \@secondoftwo
 \fi
}%
\providecommand \natexlab [1]{#1}%
\providecommand \enquote  [1]{``#1''}%
\providecommand \bibnamefont  [1]{#1}%
\providecommand \bibfnamefont [1]{#1}%
\providecommand \citenamefont [1]{#1}%
\providecommand \href@noop [0]{\@secondoftwo}%
\providecommand \href [0]{\begingroup \@sanitize@url \@href}%
\providecommand \@href[1]{\@@startlink{#1}\@@href}%
\providecommand \@@href[1]{\endgroup#1\@@endlink}%
\providecommand \@sanitize@url [0]{\catcode `\\12\catcode `\$12\catcode
  `\&12\catcode `\#12\catcode `\^12\catcode `\_12\catcode `\%12\relax}%
\providecommand \@@startlink[1]{}%
\providecommand \@@endlink[0]{}%
\providecommand \url  [0]{\begingroup\@sanitize@url \@url }%
\providecommand \@url [1]{\endgroup\@href {#1}{\urlprefix }}%
\providecommand \urlprefix  [0]{URL }%
\providecommand \Eprint [0]{\href }%
\providecommand \doibase [0]{http://dx.doi.org/}%
\providecommand \selectlanguage [0]{\@gobble}%
\providecommand \bibinfo  [0]{\@secondoftwo}%
\providecommand \bibfield  [0]{\@secondoftwo}%
\providecommand \translation [1]{[#1]}%
\providecommand \BibitemOpen [0]{}%
\providecommand \bibitemStop [0]{}%
\providecommand \bibitemNoStop [0]{.\EOS\space}%
\providecommand \EOS [0]{\spacefactor3000\relax}%
\providecommand \BibitemShut  [1]{\csname bibitem#1\endcsname}%
\let\auto@bib@innerbib\@empty
\bibitem [{\citenamefont {Dickey}(2017)}]{dickey_stretchable_2017}%
  \BibitemOpen
  \bibfield  {author} {\bibinfo {author} {\bibfnamefont {M.~D.}\ \bibnamefont
  {Dickey}},\ }\href {\doibase 10.1002/adma.201606425} {\bibfield  {journal}
  {\bibinfo  {journal} {Advanced Materials}\ }\textbf {\bibinfo {volume} {29}}
  (\bibinfo {year} {2017}),\ 10.1002/adma.201606425}\BibitemShut {NoStop}%
\bibitem [{\citenamefont {Ilievski}\ \emph {et~al.}(2011)\citenamefont
  {Ilievski}, \citenamefont {Mazzeo}, \citenamefont {Shepherd}, \citenamefont
  {Chen},\ and\ \citenamefont {Whitesides}}]{ilievski_soft_2011}%
  \BibitemOpen
  \bibfield  {author} {\bibinfo {author} {\bibfnamefont {F.}~\bibnamefont
  {Ilievski}}, \bibinfo {author} {\bibfnamefont {A.~D.}\ \bibnamefont
  {Mazzeo}}, \bibinfo {author} {\bibfnamefont {R.~F.}\ \bibnamefont
  {Shepherd}}, \bibinfo {author} {\bibfnamefont {X.}~\bibnamefont {Chen}}, \
  and\ \bibinfo {author} {\bibfnamefont {G.~M.}\ \bibnamefont {Whitesides}},\
  }\href {\doibase 10.1002/anie.201006464} {\bibfield  {journal} {\bibinfo
  {journal} {Angewandte Chemie International Edition}\ }\textbf {\bibinfo
  {volume} {50}},\ \bibinfo {pages} {1890} (\bibinfo {year}
  {2011})}\BibitemShut {NoStop}%
\bibitem [{\citenamefont {Jeong}\ \emph {et~al.}(1997)\citenamefont {Jeong},
  \citenamefont {Bae}, \citenamefont {Lee},\ and\ \citenamefont
  {Kim}}]{jeong_biodegradable_1997}%
  \BibitemOpen
  \bibfield  {author} {\bibinfo {author} {\bibfnamefont {B.}~\bibnamefont
  {Jeong}}, \bibinfo {author} {\bibfnamefont {Y.~H.}\ \bibnamefont {Bae}},
  \bibinfo {author} {\bibfnamefont {D.~S.}\ \bibnamefont {Lee}}, \ and\
  \bibinfo {author} {\bibfnamefont {S.~W.}\ \bibnamefont {Kim}},\ }\href
  {\doibase 10.1038/42218} {\bibfield  {journal} {\bibinfo  {journal} {Nature}\
  }\textbf {\bibinfo {volume} {388}},\ \bibinfo {pages} {860} (\bibinfo {year}
  {1997})}\BibitemShut {NoStop}%
\bibitem [{\citenamefont {Gong}\ \emph {et~al.}(2003)\citenamefont {Gong},
  \citenamefont {Katsuyama}, \citenamefont {Kurokawa},\ and\ \citenamefont
  {Osada}}]{gong_doublenetwork_2003}%
  \BibitemOpen
  \bibfield  {author} {\bibinfo {author} {\bibfnamefont {J.~P.}\ \bibnamefont
  {Gong}}, \bibinfo {author} {\bibfnamefont {Y.}~\bibnamefont {Katsuyama}},
  \bibinfo {author} {\bibfnamefont {T.}~\bibnamefont {Kurokawa}}, \ and\
  \bibinfo {author} {\bibfnamefont {Y.}~\bibnamefont {Osada}},\ }\href
  {\doibase 10.1002/adma.200304907} {\enquote {\bibinfo {title}
  {Double‐{Network} {Hydrogels} with {Extremely} {High} {Mechanical}
  {Strength}},}\ } (\bibinfo {year} {2003})\BibitemShut {NoStop}%
\bibitem [{\citenamefont {Sun}\ \emph {et~al.}(2012)\citenamefont {Sun},
  \citenamefont {Zhao}, \citenamefont {Illeperuma}, \citenamefont {Chaudhuri},
  \citenamefont {Oh}, \citenamefont {Mooney}, \citenamefont {Vlassak},\ and\
  \citenamefont {Suo}}]{sun_highly_2012}%
  \BibitemOpen
  \bibfield  {author} {\bibinfo {author} {\bibfnamefont {J.-Y.}\ \bibnamefont
  {Sun}}, \bibinfo {author} {\bibfnamefont {X.}~\bibnamefont {Zhao}}, \bibinfo
  {author} {\bibfnamefont {W.~R.~K.}\ \bibnamefont {Illeperuma}}, \bibinfo
  {author} {\bibfnamefont {O.}~\bibnamefont {Chaudhuri}}, \bibinfo {author}
  {\bibfnamefont {K.~H.}\ \bibnamefont {Oh}}, \bibinfo {author} {\bibfnamefont
  {D.~J.}\ \bibnamefont {Mooney}}, \bibinfo {author} {\bibfnamefont {J.~J.}\
  \bibnamefont {Vlassak}}, \ and\ \bibinfo {author} {\bibfnamefont
  {Z.}~\bibnamefont {Suo}},\ }\href {\doibase 10.1038/nature11409} {\bibfield
  {journal} {\bibinfo  {journal} {Nature}\ }\textbf {\bibinfo {volume} {489}},\
  \bibinfo {pages} {133} (\bibinfo {year} {2012})}\BibitemShut {NoStop}%
\bibitem [{\citenamefont {Liu}\ \emph {et~al.}(2017)\citenamefont {Liu},
  \citenamefont {Tan}, \citenamefont {Yu}, \citenamefont {Li}, \citenamefont
  {Abell},\ and\ \citenamefont {Scherman}}]{liu_tough_2017}%
  \BibitemOpen
  \bibfield  {author} {\bibinfo {author} {\bibfnamefont {J.}~\bibnamefont
  {Liu}}, \bibinfo {author} {\bibfnamefont {C.~S.~Y.}\ \bibnamefont {Tan}},
  \bibinfo {author} {\bibfnamefont {Z.}~\bibnamefont {Yu}}, \bibinfo {author}
  {\bibfnamefont {N.}~\bibnamefont {Li}}, \bibinfo {author} {\bibfnamefont
  {C.}~\bibnamefont {Abell}}, \ and\ \bibinfo {author} {\bibfnamefont {O.~A.}\
  \bibnamefont {Scherman}},\ }\href {\doibase 10.1002/adma.201605325}
  {\bibfield  {journal} {\bibinfo  {journal} {Adv. Mater.}\ }\textbf {\bibinfo
  {volume} {29}},\ \bibinfo {pages} {1605325} (\bibinfo {year}
  {2017})}\BibitemShut {NoStop}%
\bibitem [{\citenamefont {Tanaka}\ \emph {et~al.}(1998)\citenamefont {Tanaka},
  \citenamefont {Fukao}, \citenamefont {Miyamoto},\ and\ \citenamefont
  {Sekimoto}}]{tanaka_discontinuous_1998}%
  \BibitemOpen
  \bibfield  {author} {\bibinfo {author} {\bibfnamefont {Y.}~\bibnamefont
  {Tanaka}}, \bibinfo {author} {\bibfnamefont {K.}~\bibnamefont {Fukao}},
  \bibinfo {author} {\bibfnamefont {Y.}~\bibnamefont {Miyamoto}}, \ and\
  \bibinfo {author} {\bibfnamefont {K.}~\bibnamefont {Sekimoto}},\ }\href@noop
  {} {\bibfield  {journal} {\bibinfo  {journal} {EPL (Europhysics Letters)}\
  }\textbf {\bibinfo {volume} {43}},\ \bibinfo {pages} {664} (\bibinfo {year}
  {1998})}\BibitemShut {NoStop}%
\bibitem [{\citenamefont {Tanaka}\ \emph {et~al.}(2000)\citenamefont {Tanaka},
  \citenamefont {Fukao},\ and\ \citenamefont
  {Miyamoto}}]{tanaka_fracture_2000}%
  \BibitemOpen
  \bibfield  {author} {\bibinfo {author} {\bibfnamefont {Y.}~\bibnamefont
  {Tanaka}}, \bibinfo {author} {\bibfnamefont {K.}~\bibnamefont {Fukao}}, \
  and\ \bibinfo {author} {\bibfnamefont {Y.}~\bibnamefont {Miyamoto}},\ }\href
  {\doibase 10.1007/s101890070010} {\bibfield  {journal} {\bibinfo  {journal}
  {The European Physical Journal E}\ }\textbf {\bibinfo {volume} {3}},\
  \bibinfo {pages} {395} (\bibinfo {year} {2000})}\BibitemShut {NoStop}%
\bibitem [{\citenamefont {Livne}\ \emph {et~al.}(2005)\citenamefont {Livne},
  \citenamefont {Cohen},\ and\ \citenamefont
  {Fineberg}}]{livne_universality_2005}%
  \BibitemOpen
  \bibfield  {author} {\bibinfo {author} {\bibfnamefont {A.}~\bibnamefont
  {Livne}}, \bibinfo {author} {\bibfnamefont {G.}~\bibnamefont {Cohen}}, \ and\
  \bibinfo {author} {\bibfnamefont {J.}~\bibnamefont {Fineberg}},\ }\href
  {\doibase 10.1103/PhysRevLett.94.224301} {\bibfield  {journal} {\bibinfo
  {journal} {Physical Review Letters}\ }\textbf {\bibinfo {volume} {94}}
  (\bibinfo {year} {2005}),\ 10.1103/PhysRevLett.94.224301}\BibitemShut
  {NoStop}%
\bibitem [{\citenamefont {Livne}\ \emph {et~al.}(2008)\citenamefont {Livne},
  \citenamefont {Bouchbinder},\ and\ \citenamefont
  {Fineberg}}]{livne_breakdown_2008}%
  \BibitemOpen
  \bibfield  {author} {\bibinfo {author} {\bibfnamefont {A.}~\bibnamefont
  {Livne}}, \bibinfo {author} {\bibfnamefont {E.}~\bibnamefont {Bouchbinder}},
  \ and\ \bibinfo {author} {\bibfnamefont {J.}~\bibnamefont {Fineberg}},\
  }\href {\doibase 10.1103/PhysRevLett.101.264301} {\bibfield  {journal}
  {\bibinfo  {journal} {Physical Review Letters}\ }\textbf {\bibinfo {volume}
  {101}} (\bibinfo {year} {2008}),\ 10.1103/PhysRevLett.101.264301}\BibitemShut
  {NoStop}%
\bibitem [{\citenamefont {Livne}\ \emph {et~al.}(2010)\citenamefont {Livne},
  \citenamefont {Bouchbinder}, \citenamefont {Svetlizky},\ and\ \citenamefont
  {Fineberg}}]{livne_near-tip_2010}%
  \BibitemOpen
  \bibfield  {author} {\bibinfo {author} {\bibfnamefont {A.}~\bibnamefont
  {Livne}}, \bibinfo {author} {\bibfnamefont {E.}~\bibnamefont {Bouchbinder}},
  \bibinfo {author} {\bibfnamefont {I.}~\bibnamefont {Svetlizky}}, \ and\
  \bibinfo {author} {\bibfnamefont {J.}~\bibnamefont {Fineberg}},\ }\href
  {\doibase 10.1126/science.1180476} {\bibfield  {journal} {\bibinfo  {journal}
  {Science}\ }\textbf {\bibinfo {volume} {327}},\ \bibinfo {pages} {1359}
  (\bibinfo {year} {2010})}\BibitemShut {NoStop}%
\bibitem [{\citenamefont {Goldman}\ \emph {et~al.}(2010)\citenamefont
  {Goldman}, \citenamefont {Livne},\ and\ \citenamefont
  {Fineberg}}]{goldman_acquisition_2010}%
  \BibitemOpen
  \bibfield  {author} {\bibinfo {author} {\bibfnamefont {T.}~\bibnamefont
  {Goldman}}, \bibinfo {author} {\bibfnamefont {A.}~\bibnamefont {Livne}}, \
  and\ \bibinfo {author} {\bibfnamefont {J.}~\bibnamefont {Fineberg}},\ }\href
  {\doibase 10.1103/PhysRevLett.104.114301} {\bibfield  {journal} {\bibinfo
  {journal} {Physical Review Letters}\ }\textbf {\bibinfo {volume} {104}}
  (\bibinfo {year} {2010}),\ 10.1103/PhysRevLett.104.114301}\BibitemShut
  {NoStop}%
\bibitem [{\citenamefont {Goldman}\ \emph {et~al.}(2012)\citenamefont
  {Goldman}, \citenamefont {Harpaz}, \citenamefont {Bouchbinder},\ and\
  \citenamefont {Fineberg}}]{goldman_intrinsic_2012}%
  \BibitemOpen
  \bibfield  {author} {\bibinfo {author} {\bibfnamefont {T.}~\bibnamefont
  {Goldman}}, \bibinfo {author} {\bibfnamefont {R.}~\bibnamefont {Harpaz}},
  \bibinfo {author} {\bibfnamefont {E.}~\bibnamefont {Bouchbinder}}, \ and\
  \bibinfo {author} {\bibfnamefont {J.}~\bibnamefont {Fineberg}},\ }\href
  {\doibase 10.1103/PhysRevLett.108.104303} {\bibfield  {journal} {\bibinfo
  {journal} {Physical Review Letters}\ }\textbf {\bibinfo {volume} {108}}
  (\bibinfo {year} {2012}),\ 10.1103/PhysRevLett.108.104303}\BibitemShut
  {NoStop}%
\bibitem [{\citenamefont {Kolvin}\ \emph {et~al.}(2015)\citenamefont {Kolvin},
  \citenamefont {Cohen},\ and\ \citenamefont {Fineberg}}]{kolvin_crack_2015}%
  \BibitemOpen
  \bibfield  {author} {\bibinfo {author} {\bibfnamefont {I.}~\bibnamefont
  {Kolvin}}, \bibinfo {author} {\bibfnamefont {G.}~\bibnamefont {Cohen}}, \
  and\ \bibinfo {author} {\bibfnamefont {J.}~\bibnamefont {Fineberg}},\ }\href
  {\doibase 10.1103/PhysRevLett.114.175501} {\bibfield  {journal} {\bibinfo
  {journal} {Physical Review Letters}\ }\textbf {\bibinfo {volume} {114}}
  (\bibinfo {year} {2015}),\ 10.1103/PhysRevLett.114.175501}\BibitemShut
  {NoStop}%
\bibitem [{\citenamefont {Kolvin}\ \emph
  {et~al.}(2017{\natexlab{a}})\citenamefont {Kolvin}, \citenamefont
  {Fineberg},\ and\ \citenamefont {Adda-Bedia}}]{kolvin_nonlinear_2017}%
  \BibitemOpen
  \bibfield  {author} {\bibinfo {author} {\bibfnamefont {I.}~\bibnamefont
  {Kolvin}}, \bibinfo {author} {\bibfnamefont {J.}~\bibnamefont {Fineberg}}, \
  and\ \bibinfo {author} {\bibfnamefont {M.}~\bibnamefont {Adda-Bedia}},\
  }\href {\doibase 10.1103/PhysRevLett.119.215505} {\bibfield  {journal}
  {\bibinfo  {journal} {Phys. Rev. Lett.}\ }\textbf {\bibinfo {volume} {119}},\
  \bibinfo {pages} {215505} (\bibinfo {year} {2017}{\natexlab{a}})}\BibitemShut
  {NoStop}%
\bibitem [{\citenamefont {Kolvin}\ \emph
  {et~al.}(2017{\natexlab{b}})\citenamefont {Kolvin}, \citenamefont {Cohen},\
  and\ \citenamefont {Fineberg}}]{kolvin_topological_2017}%
  \BibitemOpen
  \bibfield  {author} {\bibinfo {author} {\bibfnamefont {I.}~\bibnamefont
  {Kolvin}}, \bibinfo {author} {\bibfnamefont {G.}~\bibnamefont {Cohen}}, \
  and\ \bibinfo {author} {\bibfnamefont {J.}~\bibnamefont {Fineberg}},\ }\href
  {\doibase 10.1038/nmat5008} {\bibfield  {journal} {\bibinfo  {journal}
  {Nature Materials}\ }\textbf {\bibinfo {volume} {17}},\ \bibinfo {pages}
  {140} (\bibinfo {year} {2017}{\natexlab{b}})}\BibitemShut {NoStop}%
\bibitem [{\citenamefont {Franck}\ \emph {et~al.}(2007)\citenamefont {Franck},
  \citenamefont {Hong}, \citenamefont {Maskarinec}, \citenamefont {Tirrell},\
  and\ \citenamefont {Ravichandran}}]{franck_three-dimensional_2007}%
  \BibitemOpen
  \bibfield  {author} {\bibinfo {author} {\bibfnamefont {C.}~\bibnamefont
  {Franck}}, \bibinfo {author} {\bibfnamefont {S.}~\bibnamefont {Hong}},
  \bibinfo {author} {\bibfnamefont {S.~A.}\ \bibnamefont {Maskarinec}},
  \bibinfo {author} {\bibfnamefont {D.~A.}\ \bibnamefont {Tirrell}}, \ and\
  \bibinfo {author} {\bibfnamefont {G.}~\bibnamefont {Ravichandran}},\ }\href
  {\doibase 10.1007/s11340-007-9037-9} {\bibfield  {journal} {\bibinfo
  {journal} {Experimental Mechanics}\ }\textbf {\bibinfo {volume} {47}},\
  \bibinfo {pages} {427} (\bibinfo {year} {2007})}\BibitemShut {NoStop}%
\bibitem [{\citenamefont {Bar-Kochba}\ \emph {et~al.}(2015)\citenamefont
  {Bar-Kochba}, \citenamefont {Toyjanova}, \citenamefont {Andrews},
  \citenamefont {Kim},\ and\ \citenamefont {Franck}}]{bar-kochba_fast_2015}%
  \BibitemOpen
  \bibfield  {author} {\bibinfo {author} {\bibfnamefont {E.}~\bibnamefont
  {Bar-Kochba}}, \bibinfo {author} {\bibfnamefont {J.}~\bibnamefont
  {Toyjanova}}, \bibinfo {author} {\bibfnamefont {E.}~\bibnamefont {Andrews}},
  \bibinfo {author} {\bibfnamefont {K.-S.}\ \bibnamefont {Kim}}, \ and\
  \bibinfo {author} {\bibfnamefont {C.}~\bibnamefont {Franck}},\ }\href
  {\doibase 10.1007/s11340-014-9874-2} {\bibfield  {journal} {\bibinfo
  {journal} {Exp Mech}\ }\textbf {\bibinfo {volume} {55}},\ \bibinfo {pages}
  {261} (\bibinfo {year} {2015})}\BibitemShut {NoStop}%
\bibitem [{\citenamefont {Steinhardt}\ and\ \citenamefont
  {Rubinstein}(2019)}]{steinhardt_rules_2019}%
  \BibitemOpen
  \bibfield  {author} {\bibinfo {author} {\bibfnamefont {W.}~\bibnamefont
  {Steinhardt}}\ and\ \bibinfo {author} {\bibfnamefont {S.}~\bibnamefont
  {Rubinstein}},\ }in\ \href@noop {} {\emph {\bibinfo {booktitle} {{APS}
  {Meeting} {Abstracts}}}}\ (\bibinfo {year} {2019})\BibitemShut {NoStop}%
\bibitem [{\citenamefont {Berg}\ \emph {et~al.}(2019)\citenamefont {Berg},
  \citenamefont {Kutra}, \citenamefont {Kroeger}, \citenamefont {Straehle},
  \citenamefont {Kausler}, \citenamefont {Haubold}, \citenamefont {Schiegg},
  \citenamefont {Ales}, \citenamefont {Beier}, \citenamefont {Rudy},
  \citenamefont {Eren}, \citenamefont {Cervantes}, \citenamefont {Xu},
  \citenamefont {Beuttenmueller}, \citenamefont {Wolny}, \citenamefont {Zhang},
  \citenamefont {Koethe}, \citenamefont {Hamprecht},\ and\ \citenamefont
  {Kreshuk}}]{berg2019}%
  \BibitemOpen
  \bibfield  {author} {\bibinfo {author} {\bibfnamefont {S.}~\bibnamefont
  {Berg}}, \bibinfo {author} {\bibfnamefont {D.}~\bibnamefont {Kutra}},
  \bibinfo {author} {\bibfnamefont {T.}~\bibnamefont {Kroeger}}, \bibinfo
  {author} {\bibfnamefont {C.~N.}\ \bibnamefont {Straehle}}, \bibinfo {author}
  {\bibfnamefont {B.~X.}\ \bibnamefont {Kausler}}, \bibinfo {author}
  {\bibfnamefont {C.}~\bibnamefont {Haubold}}, \bibinfo {author} {\bibfnamefont
  {M.}~\bibnamefont {Schiegg}}, \bibinfo {author} {\bibfnamefont
  {J.}~\bibnamefont {Ales}}, \bibinfo {author} {\bibfnamefont {T.}~\bibnamefont
  {Beier}}, \bibinfo {author} {\bibfnamefont {M.}~\bibnamefont {Rudy}},
  \bibinfo {author} {\bibfnamefont {K.}~\bibnamefont {Eren}}, \bibinfo {author}
  {\bibfnamefont {J.~I.}\ \bibnamefont {Cervantes}}, \bibinfo {author}
  {\bibfnamefont {B.}~\bibnamefont {Xu}}, \bibinfo {author} {\bibfnamefont
  {F.}~\bibnamefont {Beuttenmueller}}, \bibinfo {author} {\bibfnamefont
  {A.}~\bibnamefont {Wolny}}, \bibinfo {author} {\bibfnamefont
  {C.}~\bibnamefont {Zhang}}, \bibinfo {author} {\bibfnamefont
  {U.}~\bibnamefont {Koethe}}, \bibinfo {author} {\bibfnamefont {F.~A.}\
  \bibnamefont {Hamprecht}}, \ and\ \bibinfo {author} {\bibfnamefont
  {A.}~\bibnamefont {Kreshuk}},\ }\href {\doibase 10.1038/s41592-019-0582-9}
  {\bibfield  {journal} {\bibinfo  {journal} {Nature Methods}\ } (\bibinfo
  {year} {2019}),\ 10.1038/s41592-019-0582-9}\BibitemShut {NoStop}%
\bibitem [{\citenamefont {Menter}()}]{menter_acrylamide_nodate}%
  \BibitemOpen
  \bibfield  {author} {\bibinfo {author} {\bibfnamefont {P.}~\bibnamefont
  {Menter}},\ }\href@noop {} {\ ,\ \bibinfo {pages} {8}}\BibitemShut {NoStop}%
\bibitem [{\citenamefont {Hecht}(2012)}]{hecht2012optics}%
  \BibitemOpen
  \bibfield  {author} {\bibinfo {author} {\bibfnamefont {E.}~\bibnamefont
  {Hecht}},\ }\href@noop {} {\emph {\bibinfo {title} {Optics}}}\ (\bibinfo
  {publisher} {Pearson Education India},\ \bibinfo {year} {2012})\BibitemShut
  {NoStop}%
\bibitem [{\citenamefont {Boué}\ \emph {et~al.}(2015)\citenamefont {Boué},
  \citenamefont {Cohen},\ and\ \citenamefont {Fineberg}}]{boue_origin_2015}%
  \BibitemOpen
  \bibfield  {author} {\bibinfo {author} {\bibfnamefont {T.~G.}\ \bibnamefont
  {Boué}}, \bibinfo {author} {\bibfnamefont {G.}~\bibnamefont {Cohen}}, \ and\
  \bibinfo {author} {\bibfnamefont {J.}~\bibnamefont {Fineberg}},\ }\href
  {\doibase 10.1103/PhysRevLett.114.054301} {\bibfield  {journal} {\bibinfo
  {journal} {Physical Review Letters}\ }\textbf {\bibinfo {volume} {114}}
  (\bibinfo {year} {2015}),\ 10.1103/PhysRevLett.114.054301}\BibitemShut
  {NoStop}%
\bibitem [{kol()}]{kolvin_how_2018}%
  \BibitemOpen
  \href {\doibase 10.1103/PhysRevLett.121.135501} {\ \textbf {\bibinfo {volume}
  {121}},\ 10.1103/PhysRevLett.121.135501}\BibitemShut {NoStop}%
\bibitem [{\citenamefont {Rozen-Levy}\ \emph {et~al.}(2020)\citenamefont
  {Rozen-Levy}, \citenamefont {Kolinski}, \citenamefont {Cohen},\ and\
  \citenamefont {Fineberg}}]{rosen-levy_maximum_2020}%
  \BibitemOpen
  \bibfield  {author} {\bibinfo {author} {\bibfnamefont {L.}~\bibnamefont
  {Rozen-Levy}}, \bibinfo {author} {\bibfnamefont {J.~M.}\ \bibnamefont
  {Kolinski}}, \bibinfo {author} {\bibfnamefont {G.}~\bibnamefont {Cohen}}, \
  and\ \bibinfo {author} {\bibfnamefont {J.}~\bibnamefont {Fineberg}},\ }in\
  \href@noop {} {\emph {\bibinfo {booktitle} {{APS} {Meeting} {Abstracts}}}}\
  (\bibinfo {year} {2020})\BibitemShut {NoStop}%
\bibitem [{gol()}]{goldman_boue_failing_2015}%
  \BibitemOpen
  \href {\doibase 10.1039/C5SM00496A} {\ \textbf {\bibinfo {volume} {11}},\
  10.1039/C5SM00496A}\BibitemShut {NoStop}%
\bibitem [{\citenamefont {Freund}(1998)}]{freund_dynamic_1998}%
  \BibitemOpen
  \bibfield  {author} {\bibinfo {author} {\bibfnamefont {L.~B.}\ \bibnamefont
  {Freund}},\ }\href@noop {} {\emph {\bibinfo {title} {Dynamic {Fracture}
  {Mechanics}}}}\ (\bibinfo  {publisher} {Cambridge University Press},\
  \bibinfo {year} {1998})\BibitemShut {NoStop}%
\bibitem [{\citenamefont {Fineberg}\ and\ \citenamefont
  {Marder}(1999)}]{fineberg_instability_1999}%
  \BibitemOpen
  \bibfield  {author} {\bibinfo {author} {\bibfnamefont {J.}~\bibnamefont
  {Fineberg}}\ and\ \bibinfo {author} {\bibfnamefont {M.}~\bibnamefont
  {Marder}},\ }\href@noop {} {\bibfield  {journal} {\bibinfo  {journal}
  {Physics Reports}\ ,\ \bibinfo {pages} {108}} (\bibinfo {year}
  {1999})}\BibitemShut {NoStop}%
\end{thebibliography}%

\end{document}